# Location of upper borders of cavities containing dust and gas under pressure in comets

Sergei I. Ipatov [a,b,*]

[a] Department of Physics, Catholic University of America, Washington DC, 20064, USA
[b] Space Research Institute, Moscow, Russia

**ABSTRACT**
The distance between the pre-impact surface of Comet 9P/Tempel 1 and the upper border of the largest cavity excavated during ejection of material after the collision of the impact module of the *Deep Impact* spacecraft with the comet is estimated to be about 5-6 metres if the diameter of the DI transient crater was about 150-200 m. The estimated distance was 4 m at the diameter was 100 m. This result suggests that cavities containing dust and gas under pressure located a few metres below surfaces of comets can be frequent.

[*] Present address: Alsubai Establishment for Scientific Studies, PO Box 201536, Doha, Qatar
* E-mail: siipatov@hotmail.com

# 1 INTRODUCTION

In 2005 the impact module of the *Deep Impact* (*DI*) spacecraft collided with Comet 9P/Tempel 1 (A'Hearn et al. 2005). Ipatov & A'Hearn (2011) analysed images of the cloud of material ejected after this collision. Based on analysis of the images captured during the first 13 minutes, they studied the process of ejection of material and concluded that, in addition to the normal ejection, there was a triggered outburst of small particles. Some excess ejection was observed beginning at 1 s. The outburst was considerable 8-60 s after the impact. It increased the duration of the ejection and the mean velocities of observed ejected particles (compared with the normal ejection). The mean velocities ($\sim$100 m s$^{-1}$) of small observed particles almost were almost constant with time elapsed, $t_e$, since the impact for $10<t_e<20$ s. Ipatov & A'Hearn (2011) supposed that the outburst was caused by the ejection of material from cavities containing dust and gas under pressure. Velocities of such `fast` outburst particles could be mainly $\sim$100 m s$^{-1}$ (such velocities were obtained at various observations of the leading edge of the *DI* cloud). Hosapple & Housen (2007) supposed that ejected particles could be accelerated by the dust-gas interaction. Such acceleration could be important for time intervals of several hours. Ipatov & A'Hearn (2011) came to their conclusions based on analysis of variations in brightness in images made during the first 13 minutes and considering distances, $R$, from a place of ejection between 1 and 10 km. They concluded that particles could not increase their velocities by more than a few metres per second during not more than a few minutes, when the particles moved at $R$ between 1 and 10 km. Ipatov and A'Hearn analysed the brightness of the *DI* cloud at such distances but did not consider a plume base at $R<1$ km, in contrast to Richardson et al. (2007), who presented pictures of the evolution of the plume. In many *DI* images, the region corresponding to $R<1$ km consisted mainly of saturated pixels (in 8-bit images received from the *DI* spacecraft, the brightness of most saturated pixels was 255). Ipatov & A'Hearn (2011) analysed the sky brightness mainly at the distances $R$ at which most observed material did not fall back on the comet. According to Holsapple & Housen (2007), only 1.4 per cent of ejected *DI* material ($10^5$ kg) did not return to the comet surface.

In is difficult to explain the time variations in the brightness of the *DI* cloud at $1<R<10$ km without consideration of the triggered outburst with velocities of $\sim$100 m s$^{-1}$. Several authors (e.g. Richardson et al. 2007) assume that variation in brightness of the *DI* cloud was caused by variation in particle size distribution resulting from the striking of a layered target. Sunshine et al. (2007) concluded that sub-micron water particles were observed from 3 s to 45 min after impact and the ejection of icy particles began when the crater depth reached 1 m. We suppose that the model of a layered target plays some role in explanation of the variation of brightness of the *DI* cloud, but it cannot explain all details of such variation (for example why at $\sim$10 s there was simultaneously the jump in the direction from the place of ejection to the brightest pixel in an image of the *DI* cloud by 50$^{\circ}$, an increase in the rate of ejection of small particles, and an increase in the brightness of the brightest pixel; why at the time of ejection $t_e \sim 60$ s there was a sharp decrease in the rate of ejection of small particles, why at time $t \sim 60$ s after the impact the direction from the place of ejection to the brightest pixel returned to the direction at $1<t<12$ s, why the mean ejection velocities of observed particles were almost the same at $t_e \sim$10-20 s, etc.). Holsapple & Housen (2007) concluded that conventional cratering cannot be the sole key to observed plume of the *DI* event, and that a volatile subsurface could greatly enhance the amounts of ejected mass. Ipatov & A'Hearn (2011) noted that additional ejection at $t_e<60$ s was different for different directions. Together with variation in the direction to the brightest pixel mentioned above, this suggests of that at $t_e$~10-60 s the additional ejection was not only the result of ejection from a volatile subsurface.

Jorda et al. (2007) concluded that the diameters of particles that made contributed most to the brightness of the *DI* cloud were smaller than 3 μm. The sizes of the `fast` outburst particles could be mostly less than a few microns (although a few relatively large pieces of cavity borders could be ejected), and therefore their contribution to the total mass of ejected material would be much smaller than their contribution to the brightness of the *DI* cloud. At the beginning of the main outburst (at $t_e$≈8 s), typical velocities of ejected particles were about 100 m s$^{-1}$. Holsapple & Housen (2007) concluded that only 370 kg of ejected material would be travelling at such a velocity. As typical sizes of outburst particles were smaller than those from the normal ejection, the total mass of outburst particles ejected at that time could be much smaller than 370 kg, although their contribution to the brightness of the *DI* cloud was noticeable.

Based on studies of the ejecta plume, Richarson et al. (2007) concluded that a transient crater of not more than 85–140 m (but not less than 22–26 m) diameter formed in not more than 250–550 s. Estimates of a crater diameter made by Busko et al. (2007) on the basis of analysis of *DI* images were about 150–200 m. Schultz et al. (2007) obtained a slightly wider range for the diameter, namely 130–220 m.

On 2011 February 14, the *Stardust* spacecraft obtained photos of the crater formed after the *DI* impact. Based on analysis of the photos, Schultz, Hermalyn & Veverka (2012) concluded that two interpretations of the images are most reasonable: a large (~200 m in diameter) transient crater that collapsed, or a nested crater with a small inner pit (~50 m in diameter) surrounded by a shallow excavation crater (~180 m in diameter). Melosh and Richardson assumed (Richardson's personal communication 2011) that there is a ~50 metre diameter depression that most likely represents the *DI* crater. Schultz et al. (2012) discussed several reasons why the *Deep Impact* and *Stardust-NExT* images are inconsistent with a 50 m crater. The image of the crater made by the *Stardust* spacecraft is available on http://stardustnext.jpl.nasa.gov/Multimedia/popups/Tempel1ImageSite.html (image credit: NASA/JPL-Caltech/University of Maryland/Cornell). It is noted on this website that the crater is estimated to be 150 metres in diameter. In this image, one can see that the diameter of the brightest part of the ring zone of ejected material around the crater is about 90-100 m. The diameters of the inner and outer edges of the ring zone are ~60-70 m and ~130-140 m, respectively. The ring zone may correspond to ejected material, and the diameter of an excavation zone might be about 100 m. The crater size observed by *Stardust* spacecraft could be different from the transient crater size just after its formation. This is because the sublimation process as well as various modification processes changed the crater size after its formation.

Based on the time of the beginning of excavation of the main cavity ($t_e$≈8 s) obtained by Ipatov & A'Hearn (2011) and on the above estimates of the diameter of the *DI* crater, in Section 2 we estimate the distance $d_{cavDI}$ between the pre-impact surface of Comet Tempel 1 and the upper border of the main excavated cavity. Such estimates enabled a better understanding of the distances $d_{cav}$ between the surfaces of comets and the upper borders of cavities containing dust and gas under pressure. Possible values of $d_{cav}$ for several comets are discussed in Section 3.

## 2 CAVITIES CONTAINING DUST AND GAS UNDER PRESSURE IN COMET 9P/TEMPEL 1

It is considered (e.g. Croft 1980; Melosh 1989) that ejected material originates from an excavation cavity which has a geometry distinct from that of the transient crater. The excavation cavity and the transient crater have the same diameter $d_{tc}$, but the depth, $d_{he}$, of the excavation cavity is ~$0.1d_{tc}$, or about one-third of the transient crater depth, and, in the case of simple bowl-shaped craters, about one-half of the depth of the final apparent crater. For example, $d_{he}$~10 m for $d_{tc}$=100 m. For some craters, Croft (1980) considered the ratio $d_{he}/d_{tc}$ to be in the range [0.09, 0.17].

For theoretical models (e.g. Holsapple & Housen 2007), during the most time of crater formation (except for initial and final stages), the diameter, $d_c$, of a crater at time, $t_e$, elapsed since the impact is proportional to $t_e^\gamma$, where $\gamma$ is about 0.25-0.4. No energy dissipation corresponds to $\gamma$=0.4. Porous material has a greater dissipation; $\gamma$=0.29 for dry soils with a porosity of 30-35 per cent, and $\gamma$ ranges from 0.25 to 0.29 for highly porous materials. In their table 4, Holsapple & Housen (2007) considered models at $0.29 \leq \gamma \leq 0.355$. As comets are porous, we can probably assume that $0.25 < \gamma \leq 0.3$.

According to fig. 12 in (Holsapple 1993), the diameter, $d_c$, of a crater grows faster than $t_e^\gamma$ in the initial stage (reaching the value denoted as $d_{cav1}$), but hardly grows (and can slightly decrease) in the final stage. The duration $t_1$ of the initial stage is usually less than $0.1T_e$, where $T_e$ is the time between the beginning of ejection and the end of the intermediate stage (at this stage, $d_c$ is proportional to $t_e^\gamma$). In the mentioned figure, for water the duration $t_1$ is even less than $10^{-3}T_e$. As noted by Richardson (personal communication 2011), as a result of significant impactor penetration into the porous target, the depth of the initial excavation can grow even faster than for analytical models. The depth $d_{cavDI}$ of the DI crater at the time, $t_{eb}$, of the beginning of excavation of the main cavity is $d_{cavDI}=d_{cav1}+(d_{he}-d_{cav1})\times((t_{eb}-t_1)/(T_e-t_1))^\gamma$ for $t_1<t_{eb}<T_e$. It is difficult to estimate accurately the values of $d_{cav1}$ and $t_1$ for the DI crater.

Ipatov & A'Hearn (2011) concluded that outburst and excavation of a large cavity began ~8 s after the DI collision. Supposing $d_c$ to be proportional to $t_e^\gamma$ also at the initial stage, we can estimate the lower limit of the depth, $d_{cavDI}$, of the DI crater at the time, $t_{eb}$, of the beginning of excavation of the main cavity as $d_{cavmin}=d_{he}\times(t_{eb}/T_e)^\gamma$. At $d_{he}=0.1d_{tc}$, the value of $d_{cavmin}$ equals $d_{cmn}=0.1d_{tc}\times(t_{eb}/T_e)^\gamma$. At $t_{eb}$=8 s, the values of $d_{cmn}$ for three values of $\gamma$ are presented in Table 1. The real value of $d_{cavDI}$ is greater by some value $d_{h1}$ than $d_{cmn}$ because, during the short initial stage, the real growth of the crater is greater than that for the model used for calculation of $d_{cmn}$. We suppose that the difference $d_{h1}$ can be about 1 m. In Table 1 we consider that $d_{cavDI}=0.1d_{tc}\times(t_{eb}/T_e)^{0.3}+d_{h1}$, where $d_{h1}$=1 m.

In table 1 of Holsapple & Housen (2007), the crater formation time, $T_{cr}$, (which is slightly greater than $T_e$) is proportional to $d_{tc}$ (strength) or $d_{tc}^{1/2}$ (gravity), and $T_{cr}$=288 s at $d_{tc}$=88 m for sand-gravity scaling. At $T_{cr}$ proportional to $d_{tc}$, the latter estimate of $T_{cr}$ corresponds to 330 s at $d_{tc}$=100 m. Supposing that $T_{cr}$=330 s at $d_{tc}$=100 m (these values are also presented in table 4 in Richardson et al. 2007) and considering $T_{cr}$ to be proportional to $d_{tc}^{1/2}$ (a lower estimate) or $d_{tc}$ (an upper estimate), we obtain that $T_{cr}$ is about 470-660 s at $d_{tc}$=200 m, 400-500 s at $d_{tc}$=150 m, and 185-230 s at $d_{tc}$=50 m. For estimates of the values of $d_{cmn}$ presented in Table 1, we assumed that values of $T_e$ are equal to the upper values of the intervals of $T_{cr}$ presented above. For $T_e$=470 s and $d_{tc}$=200 m, the values of $d_{cmn}$ are greater by a factor of $2^{\gamma/2}$ ($2^{\gamma/2}$ equals 1.09, 1.11, and 1.15 at $\gamma$ equal to 0.25, 0.3, and 0.4, respectively) than the values of $d_{cmn}$ at $T_e$=660 s presented in

Table 1. Therefore, the values of $d_{cmn}$ obtained at the lower values of the above intervals of $T_{cr}$ are almost the same as those in the table.

It is hoped that the values of $T_e$ presented in Table 1 can be used. Nevertheless, below we present estimates of $d_{cavDI}$ at much smaller values of $T_e$. The smaller the values of $T_e$ are, the greater the values of $d_{cavDI}$. If the values of $T_e$ are not known, then it is possible to estimate the upper limit of $d_{cavDI}$ using the estimates of the lower limit of $T_e$. During the intermediate stage of crater formation (when the diameter $d_c$ of a crater is proportional to $t_e^\gamma$), time usually increases by more than a factor of 10 (see fig. 12 in Holsapple 1993). Therefore, during the time interval $[0.1T_e, T_e]$ $d_c$ increases by a factor of $10^\gamma$, where $10^\gamma$ equals to 2, 1.8, and 2.5 at $\gamma$ equal to 0.3, 0.25, and 0.4, respectively. These estimates show that at $t_{eb}=8$ s and $T_e>80$ s (this inequality is fulfilled for the $DI$ crater) the maximum value, $d_{cavmax}$, of $d_{cavDI}$ does not exceed $d_{he}/10^\gamma+d_{h1}$, which is in the range $[0.4d_{he}+d_{h1}, 0.56d_{he}+d_{h1}]$, that is $d_{cavmax} \leq 0.056d_{tc}+d_{h1}$ at $d_{he}/d_{tc}=0.1$ (e.g. $d_{cavmax} \leq 0.05d_{tc}+d_{h1}$ for $\gamma=0.3$). At $T_e=80$ s and $\gamma=0.3$, the value of $0.05d_{tc}$ is greater by about 2 and 5 m than the values of $d_{cmn}$ presented in Table 1 for $d_{tc}=100$ m and $d_{tc}=200$ m, respectively.

Based on the values of $d_{cavDI}$ presented in Table 1, it is concluded that the distance between the pre-impact surface of the comet and the upper border of the main excavated cavity is about 5-6 m for the estimates (150-200 m) of the diameter of the $DI$ transient crater presented by Schultz et al. (2012). The estimated distance is 4 m for the diameter of 100 m.

**Table 1.** The distance, $d_{cavDI}$, between the pre-impact surface of the comet and the upper border of the cavity excavated at $t_{eb}=8$ s for several values of the diameter, $d_{tc}$, of the $DI$ transient crater and the time, $T_e$, of formation of the crater. The minimum value, $d_{cmn}=0.1d_{tc}(t_{eb}/T_e)^\gamma$, of the distance at three values of $\gamma$ is also given. The value of $d_{cmn}$ for $\gamma=0.3$ was used for estimates of $d_{cavDI}$.

| $d_{tc}$ [m] | 50 | 100 | 150 | 200 |
|---|---|---|---|---|
| $T_e$ [s] | 230 | 330 | 500 | 660 |
| $d_{cmn}$ [m] for $\gamma=0.25$ | 2.2 | 3.9 | 5.3 | 6.6 |
| **$d_{cmn}$ [m] for $\gamma=0.3$** | 1.8 | 3.3 | 4.3 | 5.3 |
| $d_{cmn}$ [m] for $\gamma=0.4$ | 1.3 | 2.3 | 2.9 | 3.4 |
| $d_{cavDI}$ [m] | 3 | 4 | 5 | 6 |

The excavated cavity could be located at some distance from the centre of the $DI$ crater (i.e. not directly below its centre). Therefore, the distance $d_{cavDI}$ between the pre-impact surface of the comet's nucleus and the upper border of the cavity could be smaller than the depth of the crater at the beginning of excavation of the cavity. On the other hand, as a result of cracks caused by the impact, the outburst from the cavity could begin before excavation of its upper border, and consideration of cracks can increase the estimate of $d_{cavDI}$.

The largest cavity excavated after the $DI$ collision could be relatively deep because a considerable excess ejection lasted during ~50 s (at $8<t_e<60$ s). This ejection probably was from the same cavity because the direction from the place of ejection to the brightest pixel in images made at $12<t<60$ s was quite different from the direction at $t<12$ s and $t>60$ s, and one of `rays of ejection` (i.e. rays of brighter material in the DI cloud with a vertex at the place of ejection) disappeared at 60 s. The existence of `rays of ejection` in *DI* images made at $t$≈13 min argues in favour of the ejection of particles from cavities at $t_e$≈10 min. The ejection of slower-moving particles from a `fresh` surface of the *DI* crater could continue for more than 10 min.

For small cavities/cavity excavated at $t_e$≈1 s, the depth of the crater could be estimated as $d_{cmn}/8^\gamma+d_{h1}$ (where the values of $d_{cmn}$ are presented in Table 1, and $8^\gamma$ is about 2) and could be ~2-3 m.

## 3 CAVITIES IN OTHER COMETS

The distance between the pre-impact surface of Comet 9P/Tempel 1 and the upper border of the largest excavated cavity equal to about 4-6 m, and sizes of particles inside the cavities of a few microns are in good agreement with the results obtained by Kossacki & Szutowicz (2011). These authors made calculations for several models of the explosion of Comet 17P/Holmes. They concluded that the nonuniform crystallization of amorphous water ice itself is probably not sufficient for an explosion, which could be caused by a rapid sublimation of the CO ice leading to a rise of gas pressure above the tensile strength of the nucleus. In their models, the initial sublimation front of the CO ice was located at a depth of 4 m, 10 m, or 20 m, and calculations were finished when the CO pressure exceeded the threshold value 10 kPa. It was shown that the pressure of CO vapor can rise to this value only when the nucleus is composed of very fine grains, a few microns in radius.

The estimates of the locations of cavities in Comet 9P/Tempel 1 presented in Section 2 and the studies of the initial sublimation front of the CO ice in Comet 17P/Holmes discussed in the previous paragraph show that the upper borders of relatively large cavities containing dust and gas under pressure can be located at a depth of 4-20 m below surfaces of different comets. After some time, gas under pressure can make its way from a cavity to the surface of a comet, and the gas formed later will use the same way at a relatively low pressure. Therefore, probably, the more time a comet has spent close to the Sun, the greater the distances from the surface of the comet to the upper borders of cavities containing dust and gas under considerable pressure.

It is usually considered that the main sources of gas pressure are water ice sublimation, the sublimation of a more volatile ice such as CO or $CO_2$ at a lower temperature than required for water ice, and the crystallization of amorphous ice in the interior of porous nucleus. Other discussed potential mechanisms of outbursts include the polymerization of hydrogen cyanide HCN, the thermal stresses, the annealing of the amorphous water ice, and meteoritic impacts (see e.g. Gronkowski & Sacharczuk 2010; Ivanova et al. 2011; more references are given in Ipatov 2012). The porous structure of comets provides enough space for sublimation and argues in favour of existence of cavities.

The projection of the velocity of the leading edge of the *DI* cloud (onto the plane perpendicular to the line of sight) was about 100-200 m s$^{-1}$ (see references in Ipatov & A'Hearn 2011) and is typical for outburst particles ejected from comets. According to Feldman et al. (2007), in the 2005 June 14 natural outburst from Comet Tempel 1, velocities of ejection were 60-145 m s$^{-1}$. Sarugaku et al. (2010) obtained that the dust cloud caused by the outburst from Comet 217P/LINEAR expanded at a velocity of 120-140 m s$^{-1}$. Velocities of outburst particles ejected from Comet 29P/Schwassmann-Wachmann 1 were about 250±80 m s$^{-1}$ (Trigo-Rodriguez et al. 2010). Similarity of velocities of particles ejected at the triggered and natural outbursts shows that these outbursts could be caused by similar internal processes in comets.

# 4 CONCLUSIONS

The upper border of the largest cavity excavated during ejection of material after the collision of the impact module of the *Deep Impact* spacecraft with Comet 9P/Tempel 1 could be located at a depth of about 5-6 metres below the pre-impact surface of the come if the diameter of the *DI* transient crater was about 150-200 m, as suggested by Schultz et al. (2012). The estimated depth was 4 m at the diameter was 100 m. The largest cavity excavated after the *DI* collision could be relatively deep because a considerable excess ejection lasted for ~50 s.

These estimates of the depth are in accordance with the depth (4-20 m) of the initial sublimation front of the CO ice in the models of the explosion of Comet 17P/Holmes considered by Kossacki & Szutowicz (2011). Our studies suggest that cavities containing dust and gas under pressure and located a few metres below the surfaces of comets can be frequent.

## ACKNOWLEDGEMENTS

This research was supported in part by NASA through the American Astronomical Society's Small Research Grant Program.

The author is grateful to J. E. Richardson and an anonymous reviewer for useful comments.